\documentclass[sigconf]{acmart}
\AtBeginDocument{%
  }

\copyrightyear{2025}
\acmYear{2025}
\setcopyright{cc}
\setcctype{by}
\acmConference[WWW Companion '25]{Companion Proceedings of the ACM Web
Conference 2025}{April 28-May 2, 2025}{Sydney, NSW, Australia}
\acmBooktitle{Companion Proceedings of the ACM Web Conference 2025 (WWW
Companion '25), April 28-May 2, 2025, Sydney, NSW, Australia}
\acmDOI{10.1145/3701716.3715456}
\acmISBN{979-8-4007-1331-6/25/04}

\usepackage{multirow} 
\usepackage{booktabs} 
\usepackage{enumitem}
\usepackage{tabularray}

\begin{document}

\settopmatter{authorsperrow=4}

\newcommand{\method}{\textsc{HyGEN}}
\newcommand{\codelink}{https://github.com/ssong915/HyGEN}

\definecolor{bleudefrance}{rgb}{0.19, 0.55, 0.91}
\definecolor{mygreen}{rgb}{0.0, 0.5, 0.0}

\title[{\method}: Regularizing Negative Hyperedge Generation for Accurate Hyperedge Prediction]{{\method}: Regularizing Negative Hyperedge Generation \\for Accurate Hyperedge Prediction}


\author{Song Kyung Yu}
\authornotemark[1]
\email{ssong915@hanyang.ac.kr}
\affiliation{%
  \institution{Hanyang University}
  \city{Seoul}
  \country{Korea}
}

\author{Da Eun Lee}
\authornote{Both authors contributed equally to this research.}
\email{ddanable@hanyang.ac.kr}
\affiliation{%
  \institution{Hanyang University}
  \city{Seoul}
  \country{Korea}
}

\author{Yunyong Ko}
\email{yyko@cau.ac.kr}
\affiliation{%
  \institution{Chung-Ang University}
  \city{Seoul}
  \country{Korea}
}

\author{Sang-Wook Kim}
\authornote{Corresponding author.}
\email{wook@hanyang.ac.kr}
\affiliation{%
  \institution{Hanyang University}
  \city{Seoul}
  \country{Korea}
}

\renewcommand{\shortauthors}{Song Kyung Yu, Da Eun Lee, Yunyong Ko, and Sang-Wook Kim}

\begin{abstract}
  \textit{Hyperedge prediction} is a fundamental task to predict future high-order relations based on the observed network structure.
Existing hyperedge prediction methods, however, suffer from the data sparsity problem. 
To alleviate this problem, 
negative sampling methods can be used, which leverage non-existing hyperedges as contrastive information for model training. 
However, the following important challenges have been rarely studied: 
\textbf{(C1)} \textit{lack of guidance for generating negatives} and \textbf{(C2)} \textit{possibility of producing false negatives}. 
To address them, we propose a novel hyperedge prediction method, \textbf{{\method}}, that employs (1) a negative hyperedge generator that employs positive hyperedges as a guidance to generate more realistic ones and (2) a regularization term that prevents the generated hyperedges from being false negatives.
Extensive experiments on six real-world hypergraphs reveal that {\method} consistently outperforms four state-of-the-art hyperedge prediction methods. 
\end{abstract}

\ccsdesc[500]{Computing methodologies~Adversarial learning}
\ccsdesc[500]{Computing methodologies~Regularization}
\ccsdesc[500]{Computing methodologies~Network science}

\keywords{Hyperdge prediction, Adversarial learning, Negative hyperedge generation, Regularization}

\maketitle
\section{Introduction}\label{sec-intro}

In real-world networks, high-order relations (i.e., \textit{group-wise relations}) are prevalent~\cite{ko2023cash,dong2020hnhn,feng2019hgnn}, 
such as (i) a research paper co-authored by a group of researchers
and (ii) a chemical reaction co-induced by a group of proteins. 
A \textit{hypergraph}, a generalized data structure, is capable of modeling such group-wise relations as a \textit{hyperedge} without any information loss.
Due to its powerful expressiveness,
hypergraph-based network learning~\cite{feng2019hgnn,yang2022hyperle} has been widely studied and shown to outperform graph-based methods in various downstream tasks, including node classification~\cite{dong2020hnhn,choe2023whatsnet}, node ranking~\cite{zhang2021ranking, yu2021social}, and link prediction~\cite{yadati2020nhp,vaida2019link}.

\begin{figure}[t]
\centering
\includegraphics[width=0.98\linewidth]{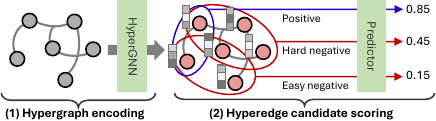}
\vspace{-3mm}
\caption{Process of hyperedge prediction.}
\vspace{-7mm}
\label{fig:prediction-process}
\end{figure}

\textit{Hyperedge prediction} (i.e., link prediction on hypergraphs) is a fundamental task in many real-world applications, such as recommender systems~\cite{zhang2021ranking} and social network analysis~\cite{yu2021social, choe2023whatsnet};
it predicts future high-order relations (i.e., hyperedges) based on an observed hypergraph structure.
A general approach to hyperedge prediction is two-fold~\cite{hwang2022ahp,ko2023cash} (see Figure~\ref{fig:prediction-process}: 
(1) (\textbf{hypergraph encoding}) the embeddings of nodes are produced by hypergraph neural networks~\cite{feng2019hgnn,dong2020hnhn} and (2) (\textbf{hyperedge candidate scoring}) the embeddings of nodes in each hyperedge candidate are \textit{aggregated} and fed into a predictor to determine whether the candidate is real.

In real-world networks, however, high-order relations are often extremely \textit{sparse}~\cite{patil2020heuristic} (i.e., $2^{|V|} \gg |E|$, where $V$ and $E$ are the sets of nodes and hyperedges, respectively).
Such a \textit{data sparsity} problem is the fundamental cause of low accuracy in hyperedge prediction.
To address this problem, negative sampling (NS) can be used~\cite{yadati2020nhp, zhang2019hyperSAGNN, hwang2022ahp}, 
utilizing non-existing hyperedges as contrastive information for model training.
Specifically, the model is trained so that positive examples get higher scores while negative examples get lower scores, 
which enhances the distinguishing ability of the model.
Thus, it is crucial to carefully choose negative hyperedges to maximize the effect of negative sampling.

However, sampling `good' negative hyperedges is challenging in the context of hyperedge prediction
since there exist too many possible negative hyperedges (i.e., $2^{|V|}-|E|$).
Although existing hyperedge prediction methods~\cite{zhang2019hyperSAGNN,yadati2020nhp,yoon2020expansion}, enhanced by negative sampling methods, have achieved breakthroughs in many fields,
they focus primarily on hypergraph encoding
while employing simple heuristic-based negative sampling methods~\cite{patil2020heuristic}.
Thus, negative sampling for hyperedge prediction is still rarely explored.

Although one recent work~\cite{hwang2022ahp} proposed an adversarial-training-based hyperedge prediction method (AHP) that leverages model-generated negative hyperedges useful for model training, it has overlooked the following two important challenges:

\vspace{0.5mm}
\noindent
\textbf{(C1) Lack of guidance for generating negatives}. 
Only a random noise signal is used as the input of the hyperedge generator in the previous method.
Thus, we posit that it may fail to effectively reflect the characteristics of positive hyperedges into generated hyperedges, 
especially in the initial stage of the training,
which could lead to inefficient and unstable training.

\vspace{0.5mm}
\noindent
\textbf{(C2) Possibility of producing false negatives}.
AHP generates negative hyperedges by using a generative adversarial network (GAN),
which aims to generate negative hyperedges similar as much as possible to positive hyperedges (i.e., copying the exact distribution of the original hyperedges).
Without any regularization for hyperedge generation, however, it might be possible to generate hyperedges too similar to positives, which would be potential positive hyperedges.

From this motivation, in this paper, 
we propose a novel adversarial-training-based method, \textbf{{\method}} which stands for regularizing negative \textbf{\underline{\textsc{Hy}}}peredge \textbf{\underline{\textsc{GEN}}}eration for accurate hyperedge prediction.
{\method} employs (1) a \textit{positive-guided negative hyperedge generator} that leverages positive hyperedges as guidance to generate more-realistic negative hyperedges for \textbf{(C1)} and (2) a \textit{regularization} term to prevent the generated hyperedges from being too similar to positive hyperedges for \textbf{(C2)}.

\vspace{1mm}
\noindent
\textbf{Contributions.} The main contributions of this work are as follows.
\begin{itemize}[leftmargin=10pt]
    \item \textbf{Challenges}: We point out two important challenges of negative sampling in hyperedge prediction: 
    \textbf{(C1)} \textit{lack of guidance for generating negatives} and \textbf{(C2)} \textit{possibility of false negatives}.
    \item \textbf{Method}: We propose a novel hyperedge prediction method, \textbf{{\method}} that employs (1) a positive-guided negative hyperedge generator for \textbf{(C1)} and (2) a regularization term for \textbf{(C2)}.
    \item \textbf{Evaluation}: Via extensive experiments on six real-world hypergraphs, we verify the superiority of {\method} over four state-of-the-art hyperedge prediction methods. 
\end{itemize}
For reproducibility, we have released the code of {\method} and datasets at: \url{\codelink}.

\section{Related Works}\label{sec-related}
\textbf{Hyperedge prediction}.
There have been a number of works to study hyperedge prediction.
they solve the hyperedge prediction problem as a classification task~\cite{yoon2020expansion,zhang2019hyperSAGNN,yadati2020nhp,hwang2022ahp}.
Expansion~\cite{yoon2020expansion} models a hypergraph as multiple \textit{n}-projected graphs and applies a logistic regression model to the projected graphs for predicting future hyperedges 
HyperSAGNN~\cite{zhang2019hyperSAGNN} uses a self-attention-based graph neural networks (GNN) model to learn hyperedges of variable sizes. 
NHP~\cite{yadati2020nhp} employs hyperedge-aware GNN models to learn node embeddings in hypergraphs, using the max-min pooling to aggregate the embeddings of nodes within each hyperedge candidate for prediction. 
AHP~\cite{hwang2022ahp}, the state-of-the-art hyperedge prediction method, employs adversarial training to generate negative hyperedges for model training and uses max-min pooling for node aggregation.

\vspace{1mm}
\noindent
\textbf{Negative hyperedge sampling}.
For enhancing the training of hyperedge prediction models, 
the following three heuristic-based methods for negative hyperedge sampling have been proposed~\cite{patil2020heuristic}:
(1) Sized NS (SNS) samples $n$ nodes uniformly at random;
(2) Motif NS (MNS) transforms a hypergraph into an ordinary graph via a clique-expansion and samples a $n$-connected component in the expanded graph;
and (3) Clique NS (CNS) selects a hyperedge $e$ and replaces one of its incident nodes $u\in e$ with a node $v\notin e$, which is linked to all the other incident nodes, i.e., ($e \setminus \{u\}) \cup \{v\}$.


\begin{table}[t]
\centering
\caption{Notations and their descriptions}
\vspace{-2mm}
\setlength\tabcolsep{1pt}
\begin{tabular}{cl}
\toprule
 \textbf{Notation} & \textbf{Description}\\
\midrule
$H$ & a hypergraph that consists of nodes and hyperedges \\
$V, E$ & the set of nodes, the set of hyperedges \\
$\mathbf{H}$ &the incidence matrix of $H$ \\
\midrule
$\mathbf{X}$ & the input node features \\
$\mathbf{P}, \mathbf{Q}$ & the node and hyperedge representations  \\
\midrule
$f(\cdot)$ & a hypergraph encoder \\
$agg(\cdot)$ & a node aggregator for hyperedge candidates \\
$pred(\cdot)$ & a hyperedge predictor \\
$G(\cdot), D(\cdot)$ & the negative hyperedge generator and discriminator \\
$enc(\cdot), dec(\cdot)$ & the encoder and decoder of the generator $G(\cdot)$ \\
\midrule
$\mathcal{L}(\cdot)$ & loss function \\
$\mathbf{W}, b$ & the learnable weight and bias matrices \\

\bottomrule
\end{tabular}
\label{table:notations}
\end{table}

\section{Proposed Method: {\method}}\label{sec-proposed}
In this section, we present a novel hyperedge prediction method, named as \textbf{{\method}}, for accurate hyperedge prediction.

\subsection{Problem Definition}
\noindent
\textbf{Notations}.
The notations used in this paper are described in Table~\ref{table:notations}.
A hypergraph is defined as \( H = (V, E) \), where \( V = \{v_1, v_2, \dots, v_{|V|}\} \) and \( E = \{e_1, e_2, \dots, e_{|E|}\} \). 
A hypergraph can generally be represented by an \textit{incidence} matrix $\mathbf{H}\in \{0,1\}^{|V|\times |E|}$,
where each element $h_{ij}=1$ if $v_i \in e_j$, and $h_{ij}=0$ otherwise.
The node and hyperedge features are represented by the matrices \( \mathbf{P} \in \mathbb{R}^{|V| \times d} \), \( \mathbf{Q} \in \mathbb{R}^{|E| \times d} \), where each row \( p_i \) and \( q_i \) represents the \( d\)-dimensional feature of a node and a hyperedge, respectively.

\vspace{1mm}
\noindent
\textbf{\textsc{Problem 1} (\textsc{Hyperedge Prediction}).} 
Given a hypergraph $\mathbf{H}\in \{0,1\}^{|V|\times |E|}$ and the initial node features $\mathbf{X}\in \mathbb{R}^{|V|\times d}$, and a hyperedge candidate $e'\notin E$,
to predict whether \( e' \) is real or not.

\subsection{Methodology}
\begin{figure}[t]
\centering
\includegraphics[width=0.95\linewidth]{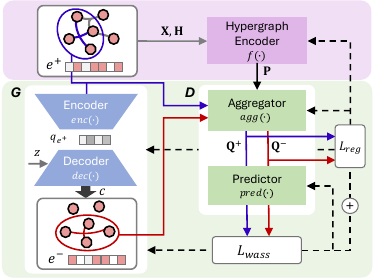}
\vspace{-3mm}
\caption{Overview of {\method}: 
(1) hypergraph encoding (\textcolor[HTML]{A973FF}{upper}) and (2) hyperedge candidate scoring (\textcolor[HTML]{71A35A}{lower}).}
\vspace{-4mm}
\label{fig:overview}
\end{figure}

\noindent
\textbf{Overview of {\method}}.
Figure~\ref{fig:overview} illustrates the overview of {\method},
which consists of (1) hypergraph encoding (\textcolor[HTML]{A973FF}{upper}) and (2) hyperedge candidate scoring (\textcolor[HTML]{71A35A}{lower}) that we focus on.

\vspace{1mm}
\noindent
\textbf{(1) Hypergraph encoding.}  
Given a hypergraph $H = (V, E)$, 
{\method} produces node embeddings $\mathbf{P} \in \mathbb{R}^{|V| \times d}$ and hyperedge embeddings $\mathbf{Q} \in \mathbb{R}^{|E| \times d}$.
Following~\cite{dong2020hnhn,chien2021allset}, {\method} adopts a \textit{2-stage aggregation} approach,
which repeats (1) (\textit{node-to-hyperedge}) producing a hyperedge embedding by aggregating the node embeddings and (2) (\textit{hyperedge-to-node}) producing a node embedding by aggregating the hyperedge embeddings.
Formally, the node and hyperedge embeddings at the \textit{l}-th layer are defined as:
\begin{align}
    \footnotesize \mathbf{Q}^{(l)} = \sigma \left(\mathbf{H}^T \mathbf{P}^{(l-1)} \mathbf{W}^{(l)}_E + b^{(l)}_E \right), 
    \footnotesize \mathbf{P}^{(l)} = \sigma \left(\mathbf{H} \mathbf{Q}^{(l)} \mathbf{W}^{(l)}_V+ b^{(l)}_V 
    \right),
    \label{eq:hypergnn}
\end{align}
where $\mathbf{P}^{(0)}=\mathbf{X}$, $\mathbf{W}^{(l)}_*$ and $b^{(l)}_*$ are trainable weight and bias matrices, respectively;
$\sigma$ is a non-linear activation function; normalization terms are omitted for simplicity in Eq.~\ref{eq:hypergnn}.

\vspace{1mm}
\noindent
\textbf{(2) Hyperedge candidate scoring.}  
The hyperedge candidate scoring of {\method} consists of a \textbf{(a) generator} to produce informative negative hyperedges for training and a \textbf{(b) discriminator} to predict whether a hyperedge candidate is positive or negative.



\vspace{1mm}
\noindent
\underline{\textbf{(2)-(a) Generator}}.
We propose a \textit{positive-guided} negative hyperedge generator for \textbf{(C1)} that employs an encoder-decoder structure.
Specifically, 
given a positive hyperedge $e^+$,
(1) the encoder produces the latent vector $q_{e^+}$ for positive hyperedge $e^+$ and
(2) the latent vector $q_{e^+}$ and a random Gaussian noise $z$ are fed into the decoder to generate a node membership vector $c \in \mathbb{R}^{|V|}$:
\begin{align}
    enc(e^+) \rightarrow q_{e^+}\in \mathbb{R}^{d},\hspace{3mm} dec(q_{e^+},z)\rightarrow c\in \mathbb{R}^{|V|}, 
    \label{eq:encoder-decoder}
\end{align}
\vspace{-1mm}
where $e^+\in \mathbb{R}^{|V|}$ is a one-hot vector whose element $e^+[i]=1$ if $i\in e^+$, and $e^+[i]=0$ otherwise,
and $c_i$ is a node membership vector whose element $c_i$ represents the probability of the node $i$ being included in the generated negative hyperedge.

For effectively extracting the characteristics of positive hyperedges and injecting them into generated negative hyperedges,
inspired by~\cite{choi2020stargan}, 
we adopt a convolutional neural network (CNN) with three layers as the architecture of our encoder and decoder.
Each of the layers consists of a 1-D convolutional layer with 256 kernels of size 3, average-pooling, and LeakyReLU as the activation function.
In the case of the decoder, 
we additionally use adaptive instance normalization (AdaIN)~\cite{huang2017adain}, which follows each convolutional layer, 
in order to inject the characteristics of a positive hyperedge into its corresponding negative hyperedge generated.

Finally, for the size $n$ of a negative hyperedge, sampled from the size distribution of positive hyperedges,
{\method} selects top-$n$ nodes as a negative hyperedge $e^-$ from the candidate probability vector $c$.
As a result, the pair of positive and negative hyperedges $e^+$ and $e^-$ are fed into the discriminator for model training.


\vspace{1mm}
\noindent
\underline{\textbf{(2)-(b) Discriminator}}. Given the learned node embeddings $\mathbf{P}$ and a hyperedge candidate $e'$ ($e^+$ or $e^-$), 
the discriminator (1) produces the embedding of a hyperedge candidate $q_{e'}$ by aggregating the embeddings of the nodes in $e'$, $\mathbf{P}[e',:]\in \mathbb{R}^{|e'| \times d}$,
and (2) computes the probability $\hat{y}_{e'}$ of $e'$ being formed based on $q_{e'}$ as:
\begin{align}
    agg(\mathbf{P}[e',:]) \rightarrow q_{e'}\in \mathbb{R}^{|d|}, \hspace{3mm} pred(q_{e'}) \rightarrow \hat{y}_{e'}\in \mathbb{R}^{1},
    \label{eq:maxpoolin}
\end{align}
where $agg(\cdot)$ is the \textit{element-wise maxmin} pooling, used as the aggregation function to reflect the diversity of the embeddings of nodes in a hyperedge candidate, by following~\cite{hwang2022ahp,yadati2020nhp},
and $pred(\cdot)$ is a hyperedge predictor, which consists of three fully-connected layers ($d \times 128 \times 8 \times 1$), followed by a sigmoid function.

\subsection{Model Training}
We train the model parameters of {\method} in an adversarial way~\cite{arjovsky2017wasserstein}:
given a batch $B$ of positive hyperedges,
(1) generate $|B|$ negative hyperedges using the generator $G$ ($enc(\cdot)$ and $dec(\cdot)$), 
(2) classify the positive and negative hyperedges using the discriminator $D$ ($agg(\cdot)$ and $pred(\cdot)$),
and (3) update the model parameters of {\method} based on their losses.
Specifically, as $D$ aims to compute the probabilities of positive hyperedges higher than those of negative hyperedges,
the loss function for $D$ is defined as:
\begin{align}
    L_D = -\frac{1}{|B|} \sum_{e^+ \in B} [D(e^+ | H, X)] + \frac{1}{|B|} \sum^{|B|}_{j=1} [D(G(z_j\mid e^+_j) | H, X)] \label{eq:loss_D},
\end{align}
where \( e^+ \) is a positive hyperedge and \( G(z_j \mid e^+_j) \) is the negative hyperedge generated from a noise \( z \) and the positive hyperedge \( e^+_j \). 
This loss $L_D$ is also used for training the hypergraph encoder $f(\cdot)$.

On the other hand, \( G \) aims to deceive \( D \) to misclassify negative hyperedges as positive.
Thus, the loss function for $G$ is defined as:
\begin{align}
    L_G = -\frac{1}{|B|} \sum^{|B|}_{j=1} [D(G(z_j\mid e^+_j) | H, X)] \label{eq:loss_G}.
\end{align}

\noindent
\underline{\textbf{Regularization for (C2)}}.
`Hard' negative hyperedges, generated by our hyperedge generator, could enhance the distinguishing ability of a hyperedge prediction model~\cite{hwang2022ahp}.
Without any regularization on a generator, however, 
it may \textit{completely copy} the distribution of the original positive hyperedges; result in generating hyperedges too similar to positive hyperedges that might potentially become positive hyperedges in the future. 
Using such hyperedges as negative hyperedges in training can lead to incorrect learning.

\begin{figure}[t]
\centering
\includegraphics[width=0.9\linewidth]{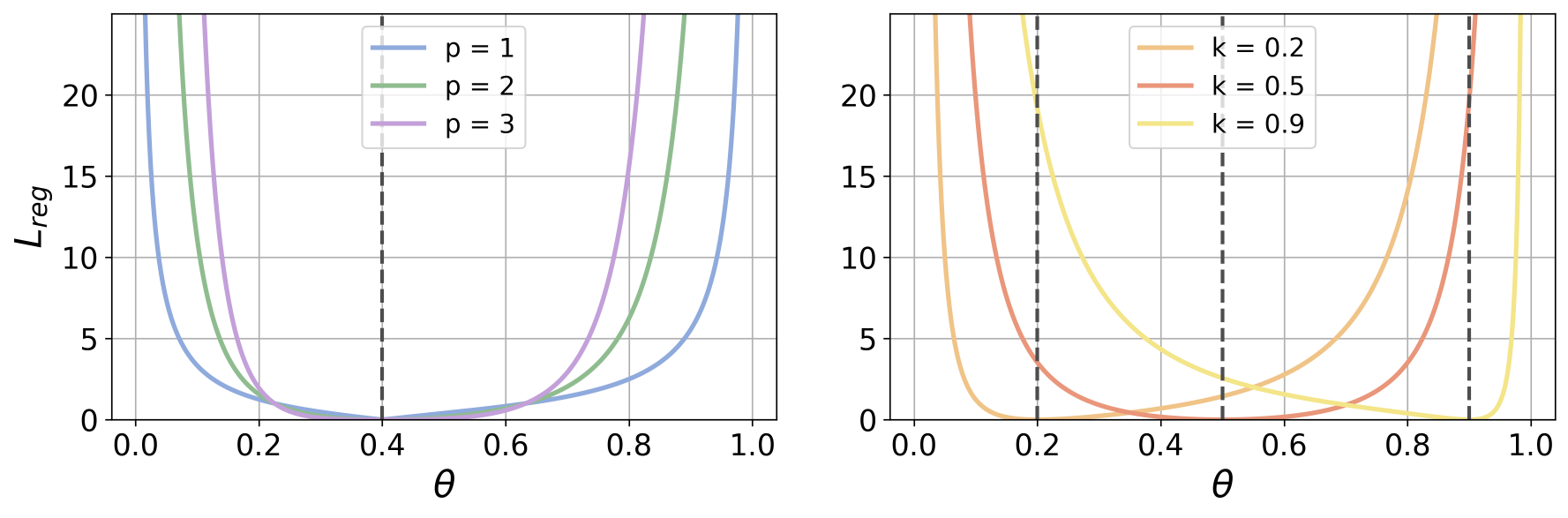}
\vspace{-5mm}
\caption{Regularization loss according to hyperparameters.}
\vspace{-4mm}
\label{fig:regularization_term}
\end{figure}
\begin{table*}[t]
\centering

\caption{Hyperedge prediction accuracy: 
{\method} always outperforms all competing methods in both the average AP and AUROC across all datasets. (The \textbf{bold} and \underline{underlined} represents the best and the second-best results in each test set, respectively.)}\label{table:eval-accuracy}
\vspace{-3mm}
\setlength\tabcolsep{3.5pt} 
\def\arraystretch{0.85} 

\scriptsize
\resizebox{\textwidth}{!}{
\begin{tabular}{c|c|rrrr|rrrr|c|rrrr|rrrr}
\toprule

\multicolumn{1}{l|}{} & \multicolumn{1}{l|}{} & \multicolumn{4}{c}{AUROC} & \multicolumn{4}{c|}{AP} & \multicolumn{1}{l|}{} & \multicolumn{4}{c}{AUROC (\%)} & \multicolumn{4}{c}{AP (\%)} \\ 

\cmidrule(lr){3-6} \cmidrule(lr){7-10} \cmidrule(lr){12-15} \cmidrule(lr){16-19} 

\multicolumn{1}{l|}{} & \multicolumn{1}{l|}{} & \multicolumn{1}{c}{SNS} & \multicolumn{1}{c}{MNS} & \multicolumn{1}{c}{CNS} & \multicolumn{1}{c}{AVG} & \multicolumn{1}{c}{SNS} & \multicolumn{1}{c}{MNS} & \multicolumn{1}{c}{CNS} & \multicolumn{1}{c|}{AVG} & \multicolumn{1}{l|}{} & \multicolumn{1}{c}{SNS} & \multicolumn{1}{c}{MNS} & \multicolumn{1}{c}{CNS} & \multicolumn{1}{c}{AVG} & \multicolumn{1}{c}{SNS} & \multicolumn{1}{c}{MNS} & \multicolumn{1}{c}{CNS} & \multicolumn{1}{c}{AVG} \\

\midrule

Expansion & \multirow{5}{*}{\rotatebox{90}{Citeseer}} & 66.3 & 78.1 & 33.1 & 59.2 & 76.5 & 81.7 & 49.8 & 69.3 & \multirow{5}{*}{\rotatebox{90}{Pubmed}} & 52.0 & 73.0 & 24.1 & 49.7 & 67.5 & 75.5 & 44.0 & 62.3 \\ 
NHP & & \textbf{99.1} & 70.1 & 51.0 & 73.4 & \textbf{99.0} & 73.1 & 52.0 & 74.7 & & \textbf{97.3} & 69.4 & 52.4 & 73.0 & \textbf{97.3} & 65.6 & 51.3 & 71.4 \\ 
HyperSAGNN & & 54.0 & 41.0 & 47.3 & 47.4 & 62.7 & 45.5 & 49.7 & 52.6 & & 52.5 & 68.6 & \textbf{54.6} & 58.6 & 53.4 & 68.0 & \underline{52.9} & 58.1 \\ 
AHP & & 94.3 & \underline{88.1} & \underline{65.1} & \underline{82.5} & 95.2 & \underline{87.0} & \underline{66.0} & \underline{82.7} & & 91.7 & \underline{84.0} & \underline{53.3} & \underline{76.3} & 91.8 & \underline{83.4} & 52.6 & \underline{75.9} \\ 
\textbf{\method} & & \underline{98.4} & \textbf{92.6} & \textbf{67.6} & \textbf{86.2} & \underline{98.5} & \textbf{91.2} & \textbf{69.4} & \textbf{86.4} & & \underline{92.1} & \textbf{87.1} & 51.6 & \textbf{77.0} & \underline{93.2} & \textbf{89.0} & \textbf{55.1} & \textbf{79.1} \\



\midrule

Expansion &  \multirow{5}{*}{\rotatebox{90}{Cora}} & 47.0 & 70.7 & 25.6 & 47.8 & 63.7 & 76.4 & 45.4 & 61.8 &  \multirow{5}{*}{\rotatebox{90}{Cora-A}} & 69.0 & 84.2 & 43.4 & 65.5 & 69.0 & 87.6 & 57.7 & 71.4 \\ 
NHP & & 94.3 & 64.1 & 47.2 & 68.5 & 94.9 & 67.8 & 50.9 & 71.2 & & 90.9 & 67.2 & 55.0 & 71.0 & 92.5 & 72.0 & 58.5 & 74.3 \\ 
HyperSAGNN & & 61.7 & 52.7 & 49.4 & 54.6 & 68.7 & 57.4 & 50.8 & 59.0 & & 38.6 & 59.1 & 54.2 & 50.6 & 53.2 & 64.3 & 54.5 & 57.3 \\ 
AHP & & \underline{96.4} & \underline{86.0} & \underline{57.2} & \underline{79.9} & \underline{96.1} & \underline{83.7} & \underline{55.2} & \underline{78.3} & & \underline{95.8} & \underline{92.4} & \underline{78.2} & \underline{88.8} & \underline{95.7} & \underline{89.8} & \underline{79.6} & \underline{88.4} \\ 

 \textbf{\method} & &  \textbf{99.1} &  \textbf{90.7} &  \textbf{58.4} &  \textbf{82.7} &  \textbf{99.0} &  \textbf{89.6} &  \textbf{57.0} &  \textbf{81.9} & &  \textbf{97.7} &  \textbf{94.7} &  \textbf{80.3} &  \textbf{90.9} &  \textbf{97.7} &  \textbf{91.4} &  \textbf{82.9} &  \textbf{90.7} \\




\midrule

Expansion &  \multirow{5}{*}{\rotatebox{90}{DBLP}} & 64.5 & 80.1 & 36.6 & 60.4 & 75.1 & \textbf{85.6} & 51.8 & 70.8 &  \multirow{5}{*}{\rotatebox{90}{DBLP-A}} & 63.4 & 82.6 & 35.0 & 60.3 & 73.0 & 85.2 & 51.2 & 69.8 \\ 

NHP & & 66.3 & 54.0 & 50.3 & 56.9 & 60.8 & 52.3 & 50.1 & 54.4 & & \textbf{96.6} & 62.3 & 55.5 & \underline{71.5} & \textbf{96.5} & 60.4 & 53.4 & 70.1 \\ 

HyperSAGNN & & 44.8 & 57.4 & 57.2 & 53.1 & 56.2 & 60.2 & \textbf{58.6} & 58.3 & & 54.8 & 79.1 & 56.3 & 63.4 & 68.6 & 80.5 & 55.2 & 68.1 \\ 

AHP & & \underline{94.6} & \underline{82.0} & \underline{56.8} & \underline{77.8} & \underline{94.7} & 81.5 & 56.1 & \underline{77.4} & & \underline{91.6} & \underline{92.6} & \underline{66.8} & \textbf{83.7} & \underline{92.8} & \underline{92.8} & \underline{70.7} & \underline{85.4} \\

 \textbf{\method} & &  \textbf{95.8} &  \textbf{83.9} &  \textbf{59.1} &  \textbf{79.6} &  \textbf{95.8} &  \underline{82.6} &  \underline{57.3} &  \textbf{78.6} & &  91.3 &  \textbf{92.7} &  \textbf{67.1} &  \textbf{83.7} &  92.7 &  \textbf{92.9} &  \textbf{71.5} &  \textbf{85.7} \\
 



\bottomrule
\end{tabular}
}
\end{table*}

To address this challenge, 
we propose a novel \textit{regularization term} that is integrated into the loss function.
This regularization loss gives a smaller penalty as the generated negative hyperedges are more similar to positive hyperedges until a certain degree and gives a larger penalty as they are too similar to positive hyperedges.
Specifically, 
given the embeddings of positive and negative hyperedges $\mathbf{Q}^{+}$ and $\mathbf{Q}^{-}$,
the regularization loss is defined as:
\begin{align}
    L_{reg} = -\left(\frac{| \theta - k |}{\theta(1-\theta)}\right)^p, \hspace{3mm}  \theta = sim(\mathbf{Q}^{+}, \mathbf{Q}^{-}) \label{eq:reg_loss}.
\end{align}
where \( k \) and \( p \) are hyperparameters that control the converge point and the curvature of the function, respectively.
Figure~\ref{fig:regularization_term} shows that regularization loss according to the hyperparameters \( k \) and \( p \).
$sim(\cdot)$ is the cosine similarity used as a similarity function.
Finally, the regularization loss is integrated into the total loss with a hyperparameter $\beta$ to control the weight of the regularization loss:
\begin{align}
    L_{total} &= L_D + L_G + \beta\cdot L_{reg} 
\end{align}
Thus, all the model parameters of {\method} are trained jointly based on the adversarial loss and the regularization loss. 

\section{Experimental Validation}\label{sec-eval}
In this section, we comprehensively evaluate {\method} by answering the following evaluation questions (EQs):
\begin{itemize}[leftmargin=10pt]
    \item \textbf{EQ1 (Accuracy)}. To what extent does {\method} improve the existing hyperedge prediction methods in terms of the accuracy?
    \item \textbf{EQ2 (Ablation study)}. Is each of our strategies beneficial to generating negative hyperedges useful for model training?   
    \item \textbf{EQ3 (Sensitivity)}. How sensitive is the effect of the regularization loss in model training to its hyperparameters ($k$ and $p$)?    
\end{itemize}

\subsection{Experimental Setups}\label{sec-eval-setup}

\noindent
\textbf{Datasets and competitors}
We use six widely used real-world hypergraphs: 
(1) three co-citation datasets (Citeseer, Cora, and Pubmed),
(2) two authorship datasets (Cora-A and DBLP-A), 
and (3) one collaboration dataset (DBLP). 
In the co-citation datasets, each node represents a paper and each hyperedge represents a group of papers co-cited by a paper;
in the authorship dataset, each node represents a paper and each hyperedge represents a group of papers written by an author;
in the collaboration dataset, each node represents a researcher and each hyperedge represents a group of researchers who wrote the same paper.
For all the datasets,
we use the bag-of-word features from the abstract of each paper as in~\cite{hwang2022ahp,ko2023cash}.
We select four state-of-the-art hyperedge prediction methods as our competitors in the experiments (Expansion~\cite{yoon2020expansion}, NHP~\cite{yadati2020nhp}, HyperSAGNN~\cite{zhang2019hyperSAGNN}, and AHP~\cite{hwang2022ahp}).

\vspace{1mm}
\noindent
\textbf{Evaluation protocol}.
We evaluate {\method} by using the protocol exactly same as that used in~\cite{hwang2022ahp}.
For each dataset, we use five data splits~\footnote{All datasets and their splits used in this paper are available at: \url{\codelink}.},
where positive hyperedges are randomly divided into training (60\%), validation (20\%), and test (20\%) sets. 
We use three validation and test sets constructed with negative hyperedges sampled by SNS, MNS, and CNS, explained in Section~\ref{sec-related}. 
As metrics, we use AUROC (area under the ROC curve) and AP (average precision).
We (1) measure AUROC and AP on each test set when the averaged AUROC over the validation sets is maximized, 
and (2) report the averaged AUROC and AP over five runs on each test set.
For all competing methods, 
we use the results reported in~\cite{hwang2022ahp} since we follow the exactly same protocol with the same data splits.

\begin{table*}[t]
\centering

\caption{Ablation study: each of our strategies is always beneficial to improving the accuracy of {\method}.}\label{table:ablation}
\vspace{-4mm}
\setlength\tabcolsep{4.25pt} 
\def\arraystretch{0.9} 

\footnotesize
\resizebox{\textwidth}{!}{
\begin{tabular}{c|cc|cc|cc|cc|cc|cc}
\toprule
\multirow{2}{*}{Method} & \multicolumn{2}{c}{Citeseer}  & \multicolumn{2}{c}{Pubmed} & \multicolumn{2}{c}{Cora} & \multicolumn{2}{c}{Cora-A} & \multicolumn{2}{c}{DBLP} & \multicolumn{2}{c}{DBLP-A} \\
\cmidrule(lr){2-3} \cmidrule(lr){4-5} \cmidrule(lr){6-7} \cmidrule(lr){8-9} \cmidrule(lr){10-11} \cmidrule(lr){12-13}

& {\scriptsize AUROC} & {\scriptsize AP} & {\scriptsize AUROC} & {\scriptsize AP} & {\scriptsize AUROC} & {\scriptsize AP} & {\scriptsize AUROC} & {\scriptsize AP} & {\scriptsize AUROC} & {\scriptsize AP} & {\scriptsize AUROC} & {\scriptsize AP} \\
\midrule
\textbf{{\method}} & \textbf{86.2} & \textbf{86.4} & \textbf{77.0} & \textbf{79.2} & \textbf{82.7} & \textbf{81.9} & \textbf{90.9} & \textbf{90.7} & \textbf{79.6} & \textbf{78.6} & \textbf{83.7} & \textbf{85.7} \\

\midrule

w/o positive-guided Gen. & \underline{85.0} \textcolor{red}{↓} & \underline{85.0} \textcolor{red}{↓} & \underline{76.9} \textcolor{red}{↓} & \underline{79.1} \textcolor{red}{↓} & \underline{80.9} \textcolor{red}{↓} & \underline{80.9} \textcolor{red}{↓} & \underline{90.2} \textcolor{red}{↓} & \underline{90.0} \textcolor{red}{↓} & \underline{76.3} \textcolor{red}{↓} & \underline{76.5} \textcolor{red}{↓} & \underline{83.0} \textcolor{red}{↓} & \underline{83.2} \textcolor{red}{↓} \\



w/o $L_{reg}$ & 82.4 \textcolor{red}{↓} & 82.9 \textcolor{red}{↓} & 76.8 \textcolor{red}{↓} & 77.1 \textcolor{red}{↓} & 78.5 \textcolor{red}{↓} & 78.2 \textcolor{red}{↓} & 89.5 \textcolor{red}{↓} & 89.4 \textcolor{red}{↓} & 72.7 \textcolor{red}{↓} & 72.4 \textcolor{red}{↓} & 70.9 \textcolor{red}{↓} & 61.6 \textcolor{red}{↓} \\

\bottomrule
\end{tabular}
}
\end{table*}

    

\begin{figure*}[t]
\centering
\includegraphics[width=0.9\linewidth]{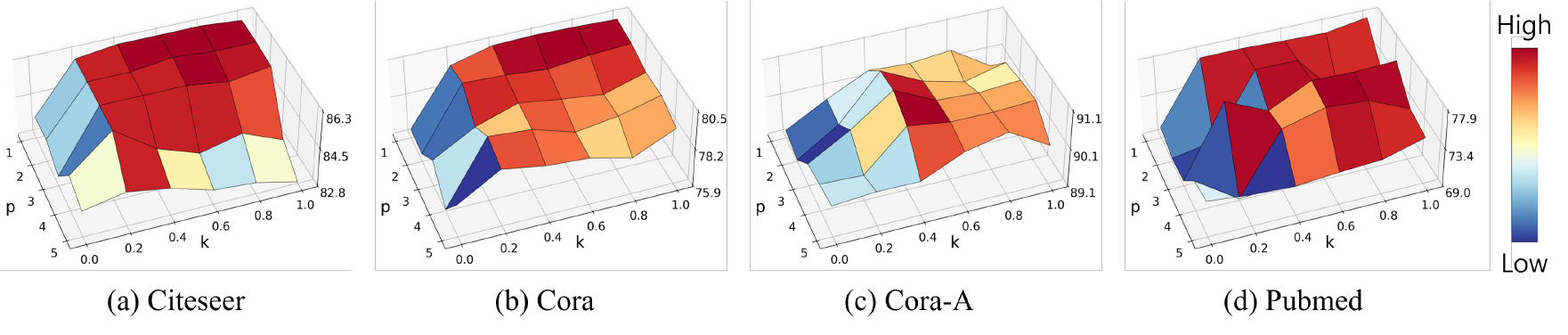}    
\vspace{-5mm}
\caption{Sensitivity analysis: {\method} achieves high accuracy across a wide range of values for the hyperparameters $k$ and $p$.}
\vspace{-4mm}
\label{fig:sensitivity-heatmap}
\end{figure*}


\subsection{EQ1. Hyperedge Prediction Accuracy}\label{sec-eval-eq1}
We first compare {\method} with four competing methods in the hyperedge prediction task.
Table~\ref{table:eval-accuracy} shows that {\method} \textit{consistently} outperforms \textit{all} competing methods in terms of \textit{both} the averaged AP and AUROC over three test sets across \textit{all} datasets.
We note that these improvements of {\method} over AHP (the best competitor) are remarkable, 
given that AHP~\cite{hwang2022ahp}, the state-of-the-art method. has already improved other existing methods significantly in those datasets. 
Via the \textit{t}-tests with a 95\% confidence level, 
we verify that the improvements of {\method} are statistically significant (i.e., the \textit{p}-values $\leq0.05$).
As a result, these results demonstrate that {\method} can generate informative negative hyperedges by effectively addressing the two challenges of negative sampling, thereby enhancing the accuracy of a hyperedge prediction task.

Interestingly, NHP~\cite{yadati2020nhp} achieves the highest accuracies in the SNS test setting of the Citeseer, Pubmed, and DBLP-A,
However, NHP shows very low accuracies on the CNS test set (i.e., the most difficult test set) of all datasets, 
which is similar to or even worse than the accuracy of the random prediction ($\approx0.5$), 
These accuracy gaps between the CNS and SNS test sets imply that NHP may be overfitting to the easy negative examples, thus which limits their ability to be generalized to other datasets. 
On the other hand, 
{\method} consistently achieves very high accuracies across all test test settings, for example 98.4\%, 92.1\%, and 91.3\% in the SNS setting of Citeseer, Pubmed, and DBLP-A, respectively. 

\subsection{EQ2. Ablation Study}\label{sec-eval-eq2}
We verify the effectiveness of our proposed strategies of {\method} individually by ablating one of them: 
(i) positive-guided negative hyperedge generator and (ii) regularization term. 
Table ~\ref{table:ablation} shows that the original version of {\method} always achieves the highest accuracy across all datasets,
which indicates that each of the proposed strategies is always beneficial to improving the accuracy of {\method}.
These results verify that our proposed strategies are able to address the two challenges successfully: \textbf{(C1)} lack of guidance for generating negatives and \textbf{(C2)} possibility of false negatives.

Furthermore, {\method} w/o $L_{reg}$ shows the worst results in all cases,
indicating that ablating the regularization term can lead to significant accuracy degradation.
These results verify that {\method} can generate negative hyperedges that are sufficiently distinct from positive hyperedges, making them useful for model training.

\subsection{EQ3. Sensitivity Analysis}\label{sec-eval-eq3}
In this experiment, we evaluate the impacts of regularization hyperparameters $k$ and $p$ on the accuracy of {\method}. 
We measure the model accuracy of {\method} with varying $k$ from 0 to 1.0 in step of 0.1 and $p$ from 1 to 5 in step of 1. 
Figure~\ref{fig:sensitivity-heatmap} shows the results,
where the $x$-axis represents the converge point hyperparameter $k$,
the $y$-axis represents the curvature hyperparameter $p$,
and the $z$-axis represents the averaged AUROC.
{\method} with $k\geq0.4$ consistently achieves higher accuracy than {\method} with $k<0.4$ regardless of $p$ (i.e., the wide red area on the surface in Figure~\ref{fig:sensitivity-heatmap}).
Based on these results, 
we believe that the accuracy of {\method} is \textit{insensitive} to the regularization hyperparameters $k$ and $p$ provided that $k\geq0.4$.

\section{Conclusion}\label{sec-con}
In this paper, we identify two key challenges of negative hyperedge sampling in the hyperedge prediction task: 
\textbf{(C1)} lack of guidance for generating negatives and \textbf{(C2)} possibility of producing false negatives. 
To address both challenges, we propose a novel hyperedge prediction method, {\method} that employs 
(1) a positive-guided negative hyperedge generator leveraging positive hyperedges as a guidance to generate informative negative hyperedges for \textbf{(C1)} and (2) a regularization term to prevent the generated hyperedges from being false negatives \textbf{(C2)}.
Comprehensive experiments on six real-world datasets verified the superiority of {\method} over four state-of-the-art hyperedge prediction methods.

\begin{acks}
This work was supported by Institute of Information \& Communications Technology Planning \& Evaluation (IITP) grant funded by the Korea government (MSIT) (RS-2022-00155586, 2022-0-00352).
\end{acks}

\bibliographystyle{ACM-Reference-Format}
\bibliography{bibliography}


\end{document}